\documentclass[journal=jacsat,manuscript=article]{achemso}
\usepackage[T1]{fontenc}
\usepackage{graphicx}
\usepackage{dcolumn}
\usepackage{bm}
\usepackage{hyperref} 
\usepackage{xcolor}
\usepackage{siunitx}
\usepackage{xspace}
\usepackage{chemformula}
\usepackage{nameref}
\usepackage{booktabs}
\usepackage{caption}
\usepackage{subcaption}
\usepackage{comment}
\hypersetup{pdfborder=0 0 0,colorlinks=true,citecolor=black,linkcolor=black}
\usepackage{amsmath}
\usepackage{amssymb}
\usepackage{nameref}
\mciteErrorOnUnknownfalse
\input epsf

\newcommand{\vecb}{\ensuremath{\mathbf}}
\newcommand\THEOSMARVEL{Theory and Simulation of Materials (THEOS), and National Centre for Computational Design and Discovery of Novel Materials (MARVEL), {\'E}cole Polytechnique F{\'e}d{\'e}rale de Lausanne, 1015 Lausanne, Switzerland}
\newcommand\PSI{Laboratory for Materials Simulations, Paul Scherrer Institut (PSI), 5232 Villigen, Switzerland}
\newcommand\BREMEN{U Bremen Excellence Chair, Bremen Center for Computational Materials 
Science, and MAPEX Center for Materials and Processes, University of Bremen, 28359 Bremen, Germany}
\newcommand\SAPIENZA{Departement of Physics, University of "La Sapienza", Rome, Italy}

\title{Extreme anharmonicity and thermal contraction of 1D wires}

\author{Chiara Cignarella}
\affiliation{\THEOSMARVEL}
\alsoaffiliation{\BREMEN}
\email{ccignare@uni-bremen.de}
\author{Lorenzo Bastonero}
\affiliation{\BREMEN}
\author{Lorenzo Monacelli}
\affiliation{\SAPIENZA}
\author{Nicola Marzari}
\affiliation{\THEOSMARVEL}
\alsoaffiliation{\PSI}
\alsoaffiliation{\BREMEN}

\date{\today}
\graphicspath{ {./figures/ause2}{./figures/cuc2}{./figures/tase3}{./figures/nequip} }

\begin{document}
\begin{abstract}
Ultrathin nanowires could play a central role in next-generation downscaled electronics. 
Here, we explore some of the most promising candidates identified from previous high-throughput screening: CuC$_2$, TaSe$_3$, and AuSe$_2$, to gain insight into the thermodynamic and anharmonic behaviors of nanowires that could be exfoliated from weakly-bonded three-dimensional materials. We analyze thermal stability, linear thermal expansion, and anharmonic heat capacity using the stochastic self-consistent harmonic approximation. Notably, our work unveils exotic features common among all the 1D wires: a colossal record negative thermal expansion and very large deviations from the Dulong-Petit law due to strong anharmonicity.
\end{abstract}


The great progress of nanotechnology has pushed the exploration of one-dimensional materials \cite{balandin2022one,meng2022one}. 
Since the first synthesis of carbon nanotubes (CNTs) \cite{iijima1991helical,tans1997individual,kim2001thermal,ouyang2002fundamental}, which, despite their unique properties, have faced difficulty in controlled synthesis, research has made significant progress, achieving the realization of ultrathin nanowires with few-atom diameters or even single-atom chains \cite{yan2017hybrid, xiao2018electrically}.
In fact, 1D materials could be highly beneficial in the continuous downscaling of next-generation electronic devices \cite{IRDS22,IRDS23}, as they naturally lack grain boundaries and edge scattering, standing out as ideal candidates for nanoelectronic applications \cite{stolyarov2016breakdown,empante2019low,zhou2015thin}.
Equally important, they represent a new playground of fundamental physical phenomena, such as Luttinger liquid, Peierls transition, and excitonic insulators \cite{peierls1955quantum, luttinger1963exactly,Varsano2017,giamarchi2016one}.
Various techniques have been explored to produce one-dimensional crystals, including self-assembly and direct growth on substrates \cite{guo2022direct,zeng2008charge,gambardella2002ferromagnetism}, directing-agents synthesis \cite{yan2017hybrid}, encapsulation inside single- or multi-walled CNTs \cite{kashtiban2021linear,stonemeyer2020stabilization}, or by chemical/mechanical exfoliation from three-dimensional crystals  \cite{stolyarov2016breakdown,island2017electronics}. The latter represents a promising direction to obtain novel wires with desired and well-defined properties.
Alongside, computational high-throughput (HT) studies can search for bulk crystals potentially exfoliable into single nanowires \cite{larsen2019definition, moustafa2022computational,campiinpreparation,cignarella2024searching}, \emph{i.e.} materials naturally composed by strongly bonded inorganic wires held together by van-der-Waals interactions, in similar fashion of what is done for two-dimensional materials \cite{cheon2017data, mounet2018two,haastrup2018computational,campi2023expansion}. 
In this context, a HT search \cite{campiinpreparation} has been conducted to discover novel one-dimensional nanowires that could be exfoliated from experimentally known and previously synthesized crystals. From the resulting database of more than 800 unique 1D, a recent work focused on metallic wires \cite{cignarella2024searching} searching for materials resilient to dynamical instabilities, such as Peierls distortions, that could represent alternatives as interconnects for future downscaled electronic devices.

In this work, we aim to gain a comprehensive understanding of the thermodynamics of exfoliable wires for nanotechnologies.
We select three wires, CuC$_2$, TaSe$_3$, and AuSe$_2$, from the previous database.
CuC$_2$ and TaSe$_3$ are two metallic wires stable at T = \SI{0}{\kelvin} \cite{cignarella2024searching}. CuC$_2$ is the thinnest metallic wire predicted to be exfoliable from 3D bulks \cite{cignarella2024searching}; it is a straight-line chain stable at 0 K when exfoliated from the bulk parent structure. The wire has the highest Young's modulus within the previous portfolio, it is bendable while maintaining the metallic state, and is robust against O$_2$ oxidation or contamination \cite{cignarella2024searching}. 
Among these materials, TaSe$_3$ was already exfoliated experimentally in 2016 \cite{stolyarov2016breakdown}, and shown to be a remarkably good candidate for downscaled applications, proposed as an alternative for local interconnects in FET \cite{liu2017low,empante2019low}.
AuSe$_2$, on the other hand, is unstable in its exfoliated metallic phase at 0 K, but it presents a stable semiconducting phase in a double-cell superstructure \cite{cignarella2024searching}. It is a candidate for the realization of Peierls transition and relevant for applications in switches or multi-functional devices, where the metal-insulator transition can be tuned by temperatures, mechanical strains or electric fields \cite{liu2016charge,zhu2018light,geremew2019bias}. 

All these materials are susceptible to charge-density-waves-like fluctuations even in the stable phase, that are typically accompanied by strong anharmonicity at finite temperature, relevant for practical applications.
The standard computational approach to deal with anharmonicity is molecular dynamics, which, however, does not include quantum effects of the nuclei. Those can play an important role, for example, in CuC$_2$ due to the presence of strong covalent bonds and light carbon atoms, leading to a Debye temperature above 1000 K.
Capturing the quantum thermodynamics of crystals beyond the harmonic approximation from first-principles requires complex simulations, such as those based on path-integral molecular dynamics \cite{ceperley1995path}. However, these techniques may become computationally very expensive, particularly when multiple materials need to be analyzed.
Recently, the stochastic self-consistent harmonic approximation (SSCHA) \cite{errea2014anharmonic,monacelli2021stochastic} has emerged as a state-of-the-art method to simulate materials beyond the harmonic approximation \cite{errea_quantum_2020, romanin2021dominant,monacelli_quantum_2023, monacelli2023first, monacelli_simulating_2024}, accounting for both quantum and thermal ionic fluctuations, while computing properties at different temperatures.
Here, for the first time, we study extensively the thermodynamics of realistic 1D materials with a full treatment of anharmonicity.
The SSCHA\cite{monacelli2021stochastic,errea2014anharmonic} minimizes the vibrational quantum free energy by optimizing a trial Gaussian ionic density matrix $\tilde{\rho}(\vecb R)$.

Nevertheless, the application of this approach to one-dimensional systems presents several challenges, arising from their inherent low-frequency phonon bands.
Real 1D systems exhibit four acoustic phonon modes instead of the three given by translational invariance: 
one-dimensional crystals are, in fact, also invariant to rotational modes around the wire axis \cite{lin2022general} - notably, this feature is not present in perfect in-line chain like carbyne \cite{romanin2021dominant}, where the rotation around the 1D axis is an identity. 
On these 1D materials, accounting for the rotational mode is essential to remove the extra fourth acoustic zero-frequency at $\Gamma$ that makes the free energy minimization extremely unstable (and leads to non-physical results).
For the purpose of this work, we lock the rotation around the wire axis, effectively implementing the acoustic sum rule (ASR) for 1D materials in SSCHA framework \cite{SSCHA_1d}.

In addition, among the four acoustic phonons, one-dimensional crystals exhibit two \textit{flexural} modes that are quadratic in the long-range limit ($\omega(q) \sim q^2)$, therefore with very low frequencies in a broad region of the Brillouin zone.
The low-frequency nature of these modes requires a high number of configurations to sample the small forces acting on them (or, alternatively, the large atomic displacements associated) \cite{aseginolaza2024bending}, which makes full ab-initio SSCHA calculations prohibitive. 
To address this limitation, we train machine-learning interatomic potentials for our systems, that allow us to compute thousands of configurations within a reasonable time and computational cost, in particular, using an equivariant neural network architecture as implemented in the open-source NequIP code \cite{batzner20223}. 
(More details can be found in the Supporting Information).

Understanding how the structure of nanowires reacts to temperature is fundamental for real applications of one-dimensional wires in downscaled electronics. Low-dimensional materials show contraction when heated \cite{mounet2005first,kwon2004thermal}. To study the behavior of nanowires upon heating, we investigate thermal expansion. That can be described by the linear expansion coefficient $\alpha$, which characterizes how the length of the wire changes with temperature. We exploit well-known thermodynamic relationships to determine the thermal expansion from simulations performed at fixed volume \cite{monacelli2023first}:
\begin{equation}\label{eq:alpha_v}
\frac{\alpha}{\beta_T} = \bigg ( \frac{\partial P}{\partial T}\bigg)_V,
\end{equation}
where $\beta_T$ is the isothermal compressibility, \emph{i.e}, the inverse of the bulk modulus, and the subscript $_V$ indicates that the derivative is performed at constant volume. 
In one-dimensional materials the volumetric thermal expansion coefficient is equivalent to the linear thermal expansion coefficient, as the expansion only occurs along the axial direction of the wire, and we use Eq. (\ref{eq:alpha_v}) with the Young's modulus Y taking the role of the bulk modulus (more details in SI).

To assess the degree of anharmonicity, we compute $\alpha$ both with the SSCHA and the quasi-harmonic approximation (QHA) \cite{ashcroft1976solid,bruesch2012phonons}.
The QHA has been successfully applied in many systems to study thermal expansion and temperature-dependent vibrational properties \cite{mounet2005first,libbi2020thermomechanical,aseginolaza2024bending}. However, QHA maintains the harmonic assumption of temperature-independent and non-interacting phonon frequencies. The role of anharmonicity is accounted for by assuming that phonon frequencies only depend on volume or, in the case of one-dimensional crystals, on the tensile strain. This approximation falls short when anharmonicity is significant, such as at high temperatures \cite{libbi2020thermomechanical}. Nevertheless, it provides a qualitative insight into thermal expansion, as $\alpha(T)$ can be decoupled into the individual contribution from each phonon mode. 

At constant volume, the heat capacity $C_V$ is defined as the derivative of the internal energy $U$ with temperature $\big( \frac{\partial U}{\partial T} \big)_V$ $\equiv$ $T \big( \frac{\partial S}{\partial T}\big)_V$, and quantifies the amount of energy required to raise the temperature of the system.
When anharmonicity is taken into account, phonon frequencies change with temperature even at constant volume, introducing an additional dependence of entropy S to temperature through phonon frequencies, that corrects the result of the harmonic theory \cite{monacelli2023first}:
\begin{equation}\label{eq:cv_sscha}
\centering
\begin{split}
    C_V 
    &  = T\frac{\partial S}{\partial T} + T \sum_{\vecb q, \mu} \frac{\partial S}{\partial \omega_{\vecb q, \mu}}\frac{\partial \omega_{\vecb q, \mu}}{\partial T}\\
    & =  \sum_{\vecb q, \mu} c_v^\text{harm}(\vecb q, \mu)\left(1+\frac{T}{\omega_{\vecb q, \mu}}\frac{\partial \omega_{\vecb q, \mu}}{\partial T}\right).
\end{split}
\end{equation}
The first term is the heat capacity of a perfect harmonic system:
\begin{equation}\label{eq:cv}
    c_v^\text{harm} (\vecb q,\mu)= k_B \bigg(\frac{\hbar \omega_{\vecb q,\mu}}{k_B T}\bigg)^2 
    \frac{
    e^{\frac{\hbar \omega_{\vecb q,\mu}}{k_B T}}
    }{\big( e^{\frac{\hbar \omega_{\vecb q,\mu}}{k_B T}}-1 \big)^2},
\end{equation}
while the second accounts for the intrinsic anharmonic changes of frequencies with temperature when the volume is fixed.
In the high-temperature limit, the harmonic contribution to the total heat capacity recovers the classical Dulong-Petit law, which becomes temperature-independent when all vibrational degrees of freedom are thermally active:
\begin{equation}
\lim_{T \to \infty} \sum_{\vecb q,\mu} c_v^\text{harm}(\vecb q,\mu) = 3Nk_B.
\label{eq:DP}
\end{equation}
This is no longer true in strongly anharmonic crystals, as phonon frequencies may change with temperature at constant-volume, even when all vibrations are thermally excited.

\begin{figure}[htbp]
   \centering
\includegraphics[width=1\textwidth]{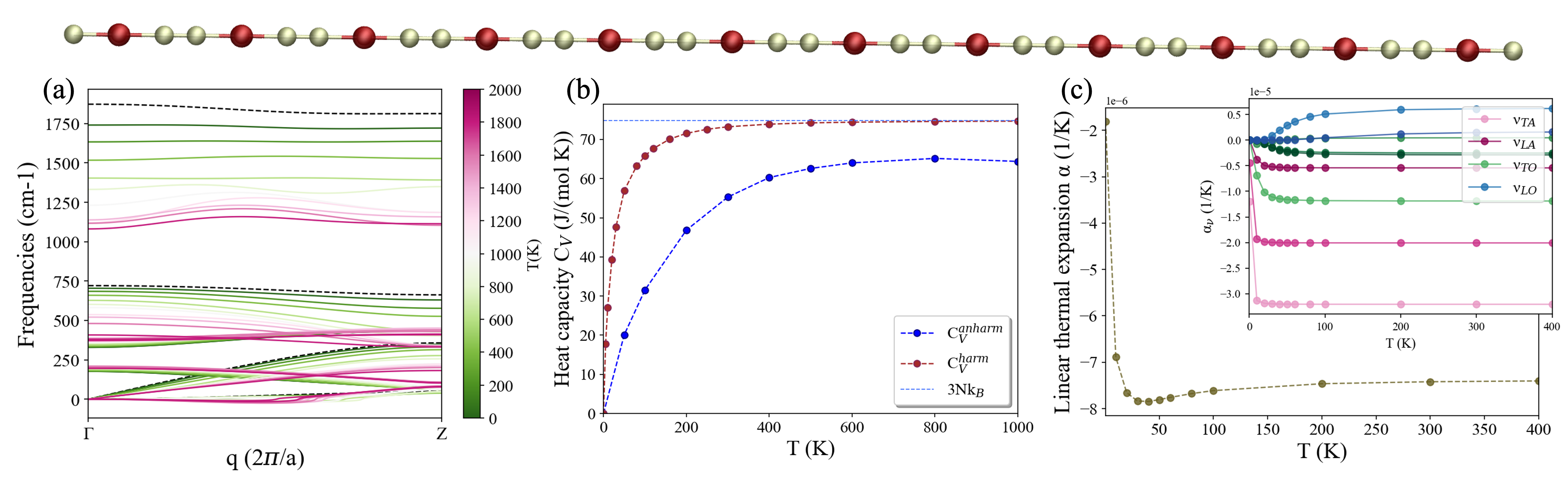}
    \caption{CuC$_2$ (Cu: red, C: gray). (a) Phonon dispersions (Hessian of the positional free energy \cite{bianco_2017_structural}) at different temperatures within the SSCHA for the wire in the configuration as exfoliated from the 3D. Dashed black lines represent harmonic phonons from DFPT. (b) Anharmonic heat capacity calculated within SSCHA with eq. (\ref{eq:cv_sscha}) (blue line) and harmonic heat capacity (brown line) using eq. (\ref{eq:cv}) with temperature-independent harmonic DFPT $\omega$. (c) Linear thermal expansion coefficient $\alpha(T)$ using QHA. In the inset, $\alpha(T)$ per phonon branch. \emph{A} and \emph{O} refer to acoustic and optical modes, \emph{T} and \emph{L} to transversal and longitudinal modes. Within the same group of modes, lighter colors correspond to lower frequencies.
    }\label{fig:cuc2}
\end{figure}

We start the discussion investigating the thermodynamic properties of CuC$_2$.
\ch{CuC2} displays strongly temperature-dependent phonon dispersions, reported in Fig. \ref{fig:cuc2}(a), which are a clear signature of anharmonicity on structural and thermodynamic properties.
To analyze the stability with temperature we define the positional free energy, which is the SSCHA free energy constraining the average ionic positions (see SI).
The calculation of the temperature-dependent anharmonic phonons from the Hessian of the positional free energy \cite{bianco_2017_structural} shows that the wire is stable in its exfoliated configuration across the entire temperature range considered, up to \SI{2000}{\kelvin}, without showing any signal of structural phase transition or emerging instabilities\footnote{All phonon frequencies are positive; note that the small wings around $\Gamma$ are due to interpolation artifacts \cite{lin2022general}.}.
Notably, the SSCHA frequencies at T = 0 K differ from the harmonic phonons, indicating a significative renormalization of phonon frequencies due to the quantum zero-point motion (ZPM) in CuC$_2$. 

In Fig. \ref{fig:cuc2}(b), we report the comparison between the specific heat at SSCHA and harmonic level: C$_V$ computed with the correct treatment of anharmonicity, using eq. (\ref{eq:cv_sscha}) (blue line), and C$_V^{harm}$ using eq. (\ref{eq:cv}) (brown line) with the harmonic frequencies at T = 0 K. 
We observe that the two quantities differ even at low-temperature, indicating a strong anharmonic contribution in this system. C$_V^{harm}$ reaches the Dulong-Petit limit (the green dashed line in the figure) at nearly 200 K, while C$_V^{anharm}$ lies far below this limit at the same temperature, and eventually never reaches it.
To further analyze the role of the anharmonicity, in SI we report and discuss on the high-temperature limit of heat capacity. We note a non-linear variation of the frequencies upon heating, entirely due to anharmonic effects, which do not saturate at the Dulong-Petit value showing instead the tendency to decrease at high-T.
This unusual feature highlights the system's intrinsic strong anharmonicity.

\begin{figure}[h!]
\centering
\includegraphics[width=0.6\textwidth]{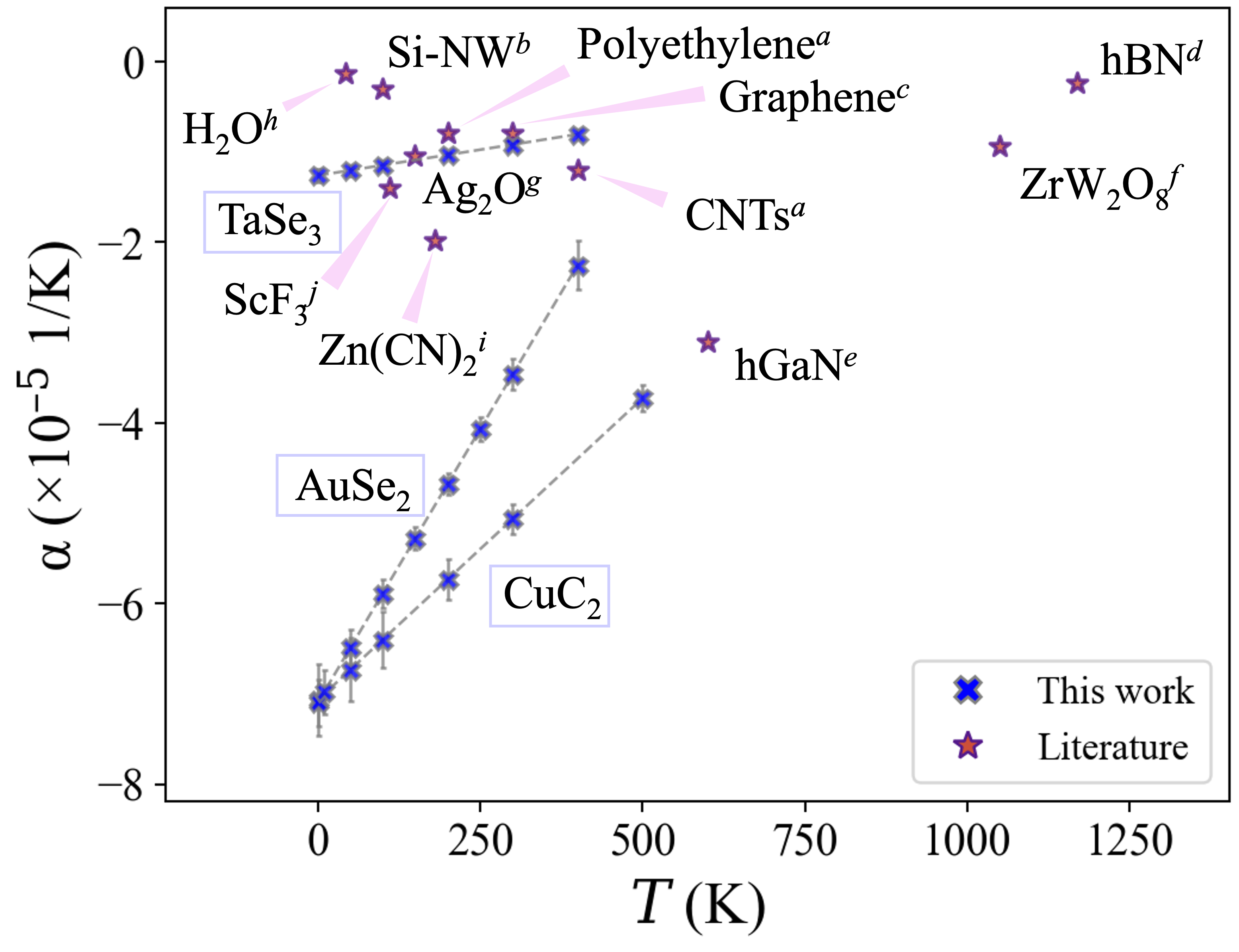}
\caption{Comparison of negative thermal expansion (NTE) coefficients for different materials: a) Ref. \citenum{kwon2004thermal}; b) Ref. \citenum{jiang2010thermal}; c) Ref. \citenum{yoon2011negative,bao2009controlled}; d) Ref. \citenum{kriegel2023incommensurability}; e) Ref. \citenum{demiroglu2021extraordinary}; f) Ref. \citenum{ernst1998phonon,mary1996negative}; g) Ref. \citenum{sanson2006negative};  h) \citenum{rottger1994lattice,fortes2018accurate}; i) Ref. \citenum{chapman2005direct}; j) Ref. \citenum{greve2010pronounced}. We plot the highest $\alpha$ value found for the material at the corresponding temperature. In blue points, the temperature-behavior of $\alpha$ for the three wires studied in this work, using SSCHA and eq. (\ref{eq:alpha_v}) (more details can be found in SI).}\label{fig:comparison}
\end{figure}

We compute the linear thermal expansion coefficient with SSCHA (see Fig. \ref{fig:comparison} and SI): $\alpha_{CuC_2}$ has negative values, with a minimum of \SI[separate-uncertainty = true]{-7.07 \pm 0.39 e-5}{\per\kelvin} - CuC$_2$ shows a contraction in its length as the temperature increases.
Thermal expansion or contraction results from the interplay between vibrational modes in the crystal.
Low-dimensional materials exhibit more freedom for out-of-plane vibrations; they can expand in the vacuum directions a relatively low energy cost due to their reduced dimensionality. 
These low-frequencies traverse modes dominate at low-temperature and can lead to an overall contraction in the longitudinal plane \cite{miller2009negative,eremenko2016role,barrera2005negative}.
However, as the temperature rises, the excitation of the high-frequency vibrations along the longitudinal axis compete with the softer ones, causing material to expand.
Negative thermal expansion (NTE) has been observed in single-walled CNTs, quasi-one-dimensional polymers, nanowires \cite{kwon2004thermal,schelling2003thermal,ho2017negative,jiang2010thermal}, or 2D monolayers such as graphene \cite{mounet2005first,libbi2020thermomechanical,yoon2011negative}.
Both graphene and CNTs exhibit a linear thermal expansion coefficient on the order of $\sim \SI{-1e-6}{\per\kelvin}$, with CNTs reaching its maximum contraction value of \SI{-1.2e-5}{\per\kelvin} at room temperature \cite{kwon2004thermal}.
The crossover occurs around \SI{2000}{\kelvin}, or \SI{900}{\kelvin}, for graphene or CNTs respectively, where $\alpha (T)$ becomes positive and the systems begin to expand.

Figure \ref{fig:cuc2}(c) reports the computed values of $\alpha(T)$ for \ch{CuC2}, using the QHA.
Like the fully anharmonic SSCHA result, also QHA yields negative in-line thermal expansion. However, the quantitative result is different due to significant role of anharmonicity, as already revealed by the analysis of the heat capacity.

While SSCHA is more accurate, QHA offers an interesting insight on the role played by individual phonon bands.
The phonon branches contributing to the negative $\alpha$ are mainly related to vibrations on the $\hat{x}$-$\hat{y}$ transversal plane (typically of the two carbon atoms), labeled \emph{TA/TO} in the inset in Fig. \ref{fig:cuc2}(c), the first two transversal acoustic modes $\nu_{TA}$ and the first optical mode $\nu_{TO}$ giving the strongest contribution. Longitudinal modes $\nu_L$ aligned with the wire axis $\hat z$ contribute positively ($\nu_{LO}$), or slighlty negatively ($\nu_{LA}$) compared to TA modes.

\begin{figure}[htbp]
   \centering
\includegraphics[width=1\textwidth]{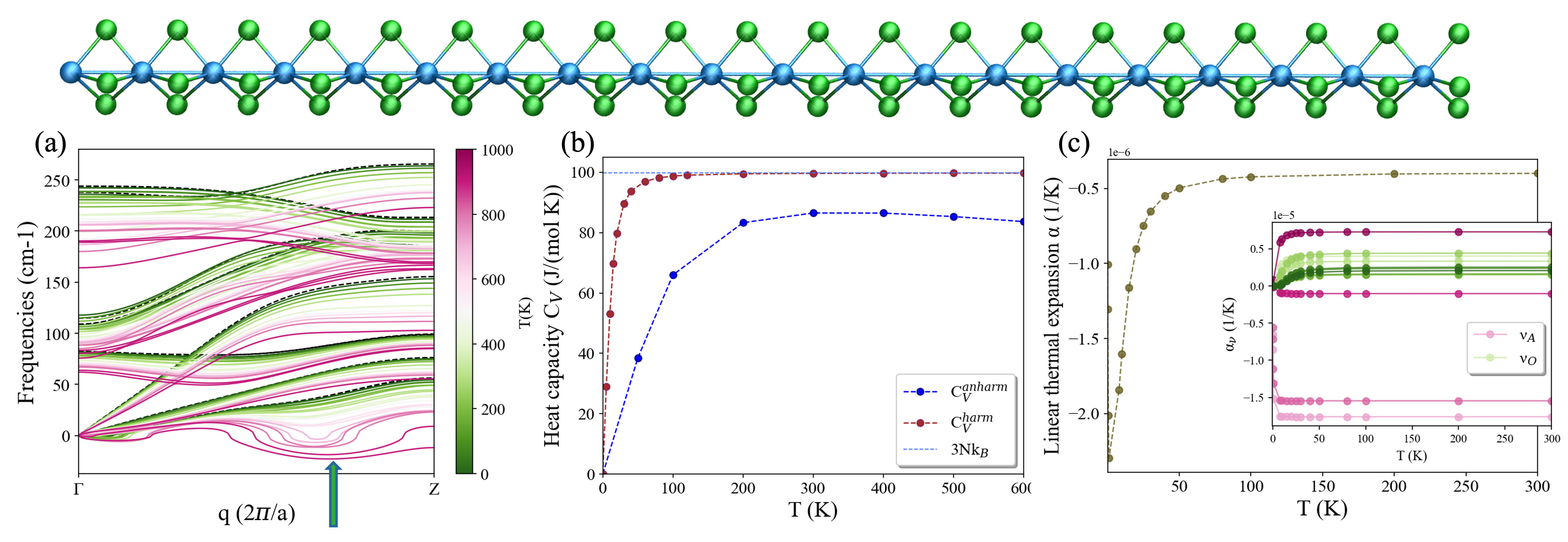}
    \caption{TaSe$_3$ (Ta: blue, Se: green). (a) Phonon dispersions (Hessian of the positional free energy) at different temperatures within the SSCHA for TaSe$_3$ in the configuration as exfoliated from the 3D\textsuperscript{1}. Dashed black lines represent harmonic phonons from DFPT. The blue arrow highlights the unstable modes appearing from \SI{700}{\kelvin}.  
    (b) Anharmonic heat capacity calculated within SSCHA with eq. (\ref{eq:cv_sscha}) (blue line) and harmonic heat capacity (brown line) using eq. (\ref{eq:cv}) with temperature-independent harmonic DFPT $\omega$. (c) Linear thermal expansion coefficient $\alpha(T)$ using QHA. In the inset, $\alpha(T)$ per phonon branch. A and O refer to acoustic and optical modes. 
    Within the same group of modes, lighter colors correspond to lower frequencies.
    \\ \textsuperscript{1}\footnotesize{Note that the small wings around $\Gamma$ are interpolation defects and sensitive to the ASR imposition, in contrast to the genuine instability at \SI{700}{\kelvin}, which aligns with an explicit q-point in the sampling mesh.}}\label{fig:tase3}
\end{figure}

Fig. \ref{fig:tase3}(a) displays the temperature-dependent phonon band structure obtained from the Hessian of the positional free energy \cite{bianco_2017_structural} of the single wire of \ch{TaSe3}. This analysis unveils the emergence of instabilities with temperature, manifesting as imaginary phonon frequencies and pointed by the blue arrow in the plot.
The system, stable at low temperature, acquires an imaginary (negative) acoustic phonon above \SI{650}{\kelvin}, marking the upper critical T above which the one-dimensional chain becomes unstable. The instability is located at $\sim$ 3/4 between $\Gamma$ and $Z$ (1,1,1/2)[2$\pi$/a] and is due to the movement of selenium atoms along the perpendicular plane to the wire axis (see SI).
This phase transition may represent a novel high-temperature charge-density wave (CDW) state of TaSe$_3$ or the thermal melting of the structure, which is likely to occur to real systems at such high temperatures.
Notably, TaSe$_3$ wire remains (meta)stable until \SI{650}{\kelvin} in the exfoliated phase from the 3D-parent compound, with remarkable perspectives for technological applications. 

In Fig. \ref{fig:tase3}(b) the comparison between harmonic and anharmonic heat capacity is shown. As before, the anharmonicity of TaSe$_3$ manifests in the strong difference between $C_V$ computed with the SSCHA and the harmonic calculations.
The high-temperature limit of the heat capacity presented in SI shows a decreasing trend with temperature, further underlining the strong anharmonicity.

The SSCHA thermal expansion coefficient computed for TaSe$_3$ results in a minimum $\alpha = \SI[separate-uncertainty = true]{-1.26(0.07)e-5}{\per\kelvin}$ (see Fig. \ref{fig:comparison} and SI). Also TaSe$_3$ exhibits contraction, smaller then CuC$_2$ yet comparable with the maximum contraction reported for CNTs \cite{kwon2004thermal}, and other well-known NTE materials.
QHA (Figure \ref{fig:tase3}(c)) shows that this is mostly given by the first two lowest-frequencies acoustic modes (lighter violet), representing symmetric vibrations of the Se atoms on the $\hat{x}$-$\hat{y}$ plane that produce a bending of the wire; this is followed by the third lowest-frequency acoustic mode, associated with twisting, whereas the longitudinal acoustic mode (darkest violet) yields instead a positive contribution.

\begin{figure}[htbp]
   \centering
\includegraphics[width=1\textwidth]{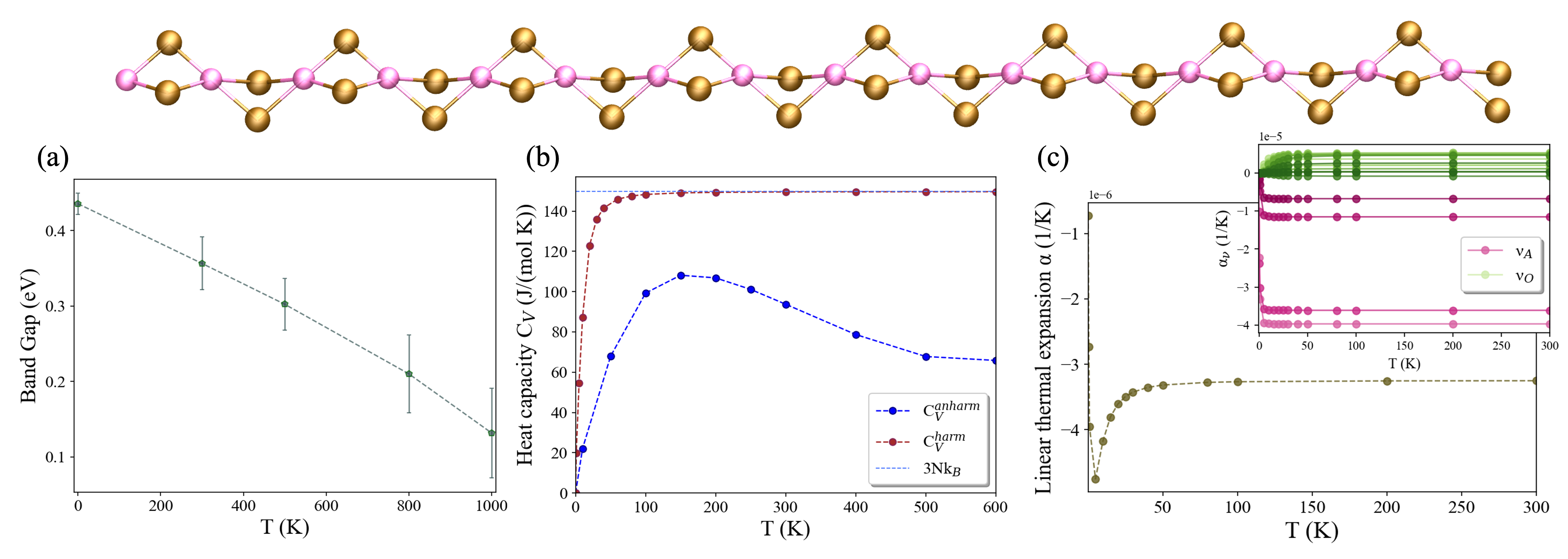}
    \caption{Au$_2$Se$_2$ (Au: pink, Se: brown). (a) Band gap behavior with temperature in the stable insulating double cell phase. (b) Anharmonic heat capacity calculated within SSCHA with eq. (\ref{eq:cv_sscha}) (blue line) and harmonic heat capacity (brown line) using eq. (\ref{eq:cv}) with temperature-independent harmonic DFPT $\omega$. (c) Linear thermal expansion coefficient $\alpha(T)$ using QHA. In the inset, $\alpha(T)$ per phonon branch. \emph{A} and \emph{O} Within the same group of modes, lighter colors correspond to lower frequencies. 
    }\label{fig:ause2}
\end{figure}

We investigate the properties of AuSe$_2$ in the reconstructed stable phase with double-cell periodicity \cite{cignarella2024searching}, which is insulating and with a Kohn-Sham band gap at the PBE level E$_g$=0.48 eV \cite{cignarella2024searching}. 
We inspect the behavior of the band gap with temperature\cite{giustino2017electron,monacelli2021black}, illustrated in Fig. \ref{fig:ause2}(a): notably, E$_g$ decreases significantly, from 0.48 eV at 0 K (0.43 $\pm$ 0.01 eV due to ZPM) to 0.13 $\pm$ 0.06 eV at 1000 K. This temperature-dependent band gap can be exploited, for instance, in optoelectronic devices or temperature sensors \cite{wu2003temperature,choi2017temperature}.

We report the anharmonic heat capacity of AuSe$_2$, Fig. \ref{fig:ause2}(b), and its comparison with the 3Nk$_B$ limit: same as the previous cases, AuSe$_2$ reveals considerable anharmonic, with strong difference between C$_V^{harm}$ and C$_V^{anharm}$ already at small temperatures.
The high-temperature limit of the heat capacity (see SI) shows a similar decreasing trend to TaSe$_3$ with a non-linearity $\omega$ variation even stronger than CuC$_2$, which makes the heat capacity to diverge from the Dulong-Petit law.

The AuSe$_2$ minimum linear thermal coefficient using SSCHA is $\alpha =\SI[separate-uncertainty = true]{-7.10(0.25)e-5}{\per\kelvin}$ (see Fig. \ref{fig:comparison} and SI). This value, as in CuC$_2$, outlines a giant thermal contraction, superior to all other materials showing NTE.
In Fig. \ref{fig:ause2}(c), the QHA results for $\alpha$ are presented. The temperature dependence resembles that observed in other materials such as graphene and CNTs \cite{mounet2005first,kwon2004thermal}, but here we find a much sharper decrease, with the maximal contraction occurring at extremely low temperatures.
As before, the QHA results give just a qualitative understanding of this trend. 
We recognise that the four acoustic phonons are responsible for the contraction (visible in the inset), while optical phonons instead present smaller and positive $\alpha$. 
All the acoustic eigenmodes $\nu_A$ corresponds to Se-oscillations along $\hat x$ and $\hat y$, causing bending or twisting of the wire, with minor movements of Au along the same transversal direction.




In summary, we investigated the thermodynamic properties of one-dimensional wires employing the SSCHA to analyze thermal stability, heat capacity and thermal expansion, fully accounting for quantum and anharmonic effects. 
Our results uncover thermodynamic behaviors that are common across all the three materials studied, despite their diverse characteristic (a metal, a high-temperature unstable system, and an insulators with CDW), suggesting that our findings are general features of 1D materials, further supported by similar observations in carbon nanotubes and cumulene.

We observe extremely strong anharmonic behavior, underscoring the importance of anharmonicity in the thermodynamics of 1D systems.

All the three wires exhibit an extremely large negative thermal expansion.
Materials that contract with temperature are relatively rare but extremely useful in applications, for example, to compensate the thermal expansion in composite materials. NTE has been observed in quasi-1D-polymers, CNTs or nanowires \cite{ho2017negative,kwon2004thermal,jiang2010thermal,jiang2004thermal}, in layered materials such as graphene and 2D nitrides \cite{bao2009controlled,mounet2005first,yoon2011negative,kriegel2023incommensurability,demiroglu2021extraordinary} and in three-dimensional crystals like cubic ZrW$_2$O$_8$ and related compounds, ZrV$_2$O$_7$ family, ScF$_3$, and cuprite structures \cite{ernst1998phonon,mary1996negative,sanson2006negative,greve2010pronounced,barrera2005negative}.
We also mention that the 3D compound Ag$_3$[Co(CN)$_6$]\cite{goodwin2008colossal} exhibits an anisotropic thermal expansion coefficient $\alpha_c$ of -12 $\times 10^{-5}$ $K^{-1}$ along the $c$-axis, while positive thermal expansion along the other two axes (in contrast to the isotropic contraction observed in previously mentioned NTE materials, and resulting in an overall positive thermal expansion). 
A comparison of $\alpha$ coefficients for different known NTE materials is presented in Fig. \ref{fig:comparison}. As visible, CuC$_2$ and AuSe$_2$ stands out as the greatest contraction among all. 


We noted that in all the three cases, the contraction is due to softening of vibrational modes (mostly acoustic) along the $\hat x$-$\hat y$ perpendicular directions of the wire, typically twisting and bending of the wires.
This behavior is observed in other low-dimensional systems \cite{mounet2005first,kwon2004thermal}, and has been studied in variuos 2D materials \cite{bao2009controlled,aseginolaza2024bending,kriegel2023incommensurability}. The modulation of the monolayer on the transversal directions gives rise to ripples \cite{bao2009controlled} predicted by Liftshitz in 1952 \cite{lifshitz1952thermal} and known as \emph{membrane effect} \cite{mounet2005first}.

In conclusion, one-dimensional materials offer significant promise for technological applications, and the three materials analyzed here are a notable example. This study demonstrates that all the wires analyzed show remarkable thermodynamic properties. A systematic exploration that rigorously accounts for anharmonicity could uncover additional outstanding candidates for future applications.


\section{Supporting Information}
Detailed computational methods, with parameters employed in the calculations. Details on the calculations of the linear thermal expansion coefficient, with component of the SSCHA stress-tensor along the wire in function of temperature for the three materials, together to a discussion on error sources. 
High-temperature limit of the specific heat capacity for CuC$_2$, TaSe$_3$, and AuSe$_2$ with discussion. 
TaSe$_3$ unstable structures following the unstable modes at high temperature.
Details on the training of the machine learning potentials, with discussion of ML errors and comparison of harmonic phonon dispersions between ML and explicit DFPT calculations.
\subsection{Code and data availability}
The implemented ASR for 1D materials are made available on GitHub as part of the \texttt{python-sscha} distribution, Ref. \citenum{SSCHA_1d}. 
The AiiDA interface used to handle the explicit DFT calculations is available through the \texttt{AiiDAEnsemble} class, from SSCHA version 1.4.0 \cite{SSCHA_github}. 
The trained force fields for the three wires are freely available and can be found on the Materials Cloud archive \cite{cignarella2025cloud}, together with all the relevant input and output files \cite{cignarella2025cloud}.
\section{Acknowledgments}
This research was supported by the NCCR MARVEL, a National Centre of Competence in Research, funded by the Swiss National Science Foundation (grant number 205602).
L.B. and N.M. acknowledge support from the Deutsche Forschungsgemeinschaft (DFG) under Germany’s Excellence Strategy (EXC 2077, No. 390741603, University Allowance, University of Bremen).
The authors acknowledge the Swiss National Supercomputing Centre (CSCS) for simulation time under project ID mr33. 
We thank D. Campi, T. Chiarotti and F. Libbi for fruitful discussions.

\textbf{Conflict of Interest}: The authors declare no competing financial interest

\bibliography{apssamp}

\providecommand{\latin}[1]{#1}
\makeatletter
\providecommand{\doi}
  {\begingroup\let\do\@makeother\dospecials
  \catcode`\{=1 \catcode`\}=2 \doi@aux}
\providecommand{\doi@aux}[1]{\endgroup\texttt{#1}}
\makeatother
\providecommand*\mcitethebibliography{\thebibliography}
\csname @ifundefined\endcsname{endmcitethebibliography}  {\let\endmcitethebibliography\endthebibliography}{}
\begin{mcitethebibliography}{80}
\providecommand*\natexlab[1]{#1}
\providecommand*\mciteSetBstSublistMode[1]{}
\providecommand*\mciteSetBstMaxWidthForm[2]{}
\providecommand*\mciteBstWouldAddEndPuncttrue
  {\def\EndOfBibitem{\unskip.}}
\providecommand*\mciteBstWouldAddEndPunctfalse
  {\let\EndOfBibitem\relax}
\providecommand*\mciteSetBstMidEndSepPunct[3]{}
\providecommand*\mciteSetBstSublistLabelBeginEnd[3]{}
\providecommand*\EndOfBibitem{}
\mciteSetBstSublistMode{f}
\mciteSetBstMaxWidthForm{subitem}{(\alph{mcitesubitemcount})}
\mciteSetBstSublistLabelBeginEnd
  {\mcitemaxwidthsubitemform\space}
  {\relax}
  {\relax}

\bibitem[Balandin \latin{et~al.}(2022)Balandin, Kargar, Salguero, and Lake]{balandin2022one}
Balandin,~A.~A.; Kargar,~F.; Salguero,~T.~T.; Lake,~R.~K. {One-dimensional van der Waals quantum materials}. \emph{Materials Today} \textbf{2022}, \emph{55}, 74--91\relax
\mciteBstWouldAddEndPuncttrue
\mciteSetBstMidEndSepPunct{\mcitedefaultmidpunct}
{\mcitedefaultendpunct}{\mcitedefaultseppunct}\relax
\EndOfBibitem
\bibitem[Meng \latin{et~al.}(2022)Meng, Wang, and Ho]{meng2022one}
Meng,~Y.; Wang,~W.; Ho,~J.~C. One-dimensional atomic chains for ultimate-scaled electronics. \emph{ACS Nano} \textbf{2022}, \emph{16}, 13314--13322\relax
\mciteBstWouldAddEndPuncttrue
\mciteSetBstMidEndSepPunct{\mcitedefaultmidpunct}
{\mcitedefaultendpunct}{\mcitedefaultseppunct}\relax
\EndOfBibitem
\bibitem[Iijima(1991)]{iijima1991helical}
Iijima,~S. Helical microtubules of graphitic carbon. \emph{Nature} \textbf{1991}, \emph{354}, 56--58\relax
\mciteBstWouldAddEndPuncttrue
\mciteSetBstMidEndSepPunct{\mcitedefaultmidpunct}
{\mcitedefaultendpunct}{\mcitedefaultseppunct}\relax
\EndOfBibitem
\bibitem[Tans \latin{et~al.}(1997)Tans, Devoret, Dai, Thess, Smalley, Geerligs, and Dekker]{tans1997individual}
Tans,~S.~J.; Devoret,~M.~H.; Dai,~H.; Thess,~A.; Smalley,~R.~E.; Geerligs,~L.; Dekker,~C. Individual single-wall carbon nanotubes as quantum wires. \emph{Nature} \textbf{1997}, \emph{386}, 474--477\relax
\mciteBstWouldAddEndPuncttrue
\mciteSetBstMidEndSepPunct{\mcitedefaultmidpunct}
{\mcitedefaultendpunct}{\mcitedefaultseppunct}\relax
\EndOfBibitem
\bibitem[Kim \latin{et~al.}(2001)Kim, Shi, Majumdar, and McEuen]{kim2001thermal}
Kim,~P.; Shi,~L.; Majumdar,~A.; McEuen,~P.~L. Thermal transport measurements of individual multiwalled nanotubes. \emph{Physical Review Letters} \textbf{2001}, \emph{87}, 215502\relax
\mciteBstWouldAddEndPuncttrue
\mciteSetBstMidEndSepPunct{\mcitedefaultmidpunct}
{\mcitedefaultendpunct}{\mcitedefaultseppunct}\relax
\EndOfBibitem
\bibitem[Ouyang \latin{et~al.}(2002)Ouyang, Huang, and Lieber]{ouyang2002fundamental}
Ouyang,~M.; Huang,~J.-L.; Lieber,~C.~M. Fundamental electronic properties and applications of single-walled carbon nanotubes. \emph{Accounts of Chemical Research} \textbf{2002}, \emph{35}, 1018--1025\relax
\mciteBstWouldAddEndPuncttrue
\mciteSetBstMidEndSepPunct{\mcitedefaultmidpunct}
{\mcitedefaultendpunct}{\mcitedefaultseppunct}\relax
\EndOfBibitem
\bibitem[Yan \latin{et~al.}(2017)Yan, Hohman, Li, Jia, Solis-Ibarra, Wu, Dahl, Carlson, Tkachenko, Fokin, \latin{et~al.} others]{yan2017hybrid}
Yan,~H.; Hohman,~J.~N.; Li,~F.~H.; Jia,~C.; Solis-Ibarra,~D.; Wu,~B.; Dahl,~J.~E.; Carlson,~R.~M.; Tkachenko,~B.~A.; Fokin,~A.~A.; others Hybrid metal--organic chalcogenide nanowires with electrically conductive inorganic core through diamondoid-directed assembly. \emph{Nature Materials} \textbf{2017}, \emph{16}, 349--355\relax
\mciteBstWouldAddEndPuncttrue
\mciteSetBstMidEndSepPunct{\mcitedefaultmidpunct}
{\mcitedefaultendpunct}{\mcitedefaultseppunct}\relax
\EndOfBibitem
\bibitem[Xiao \latin{et~al.}(2018)Xiao, Burg, Zhou, Yan, Wang, Ding, Reed, Miller, and Dauskardt]{xiao2018electrically}
Xiao,~Q.; Burg,~J.~A.; Zhou,~Y.; Yan,~H.; Wang,~C.; Ding,~Y.; Reed,~E.; Miller,~R.~D.; Dauskardt,~R.~H. Electrically Conductive Copper Core--Shell Nanowires through Benzenethiol-Directed Assembly. \emph{Nano Letters} \textbf{2018}, \emph{18}, 4900--4907\relax
\mciteBstWouldAddEndPuncttrue
\mciteSetBstMidEndSepPunct{\mcitedefaultmidpunct}
{\mcitedefaultendpunct}{\mcitedefaultseppunct}\relax
\EndOfBibitem
\bibitem[{IEEE International Roadmap for Devices and Systems}(2022)]{IRDS22}
{IEEE International Roadmap for Devices and Systems} {Executive Summary 2022}. \url{https://irds.ieee.org/}, 2022; Institute of Electrical and Electronics Engineers\relax
\mciteBstWouldAddEndPuncttrue
\mciteSetBstMidEndSepPunct{\mcitedefaultmidpunct}
{\mcitedefaultendpunct}{\mcitedefaultseppunct}\relax
\EndOfBibitem
\bibitem[{IEEE International Roadmap for Devices and Systems}(2023)]{IRDS23}
{IEEE International Roadmap for Devices and Systems} {2023 Chairman's Editorial}. \url{https://irds.ieee.org/}, 2023; P. Gargini\relax
\mciteBstWouldAddEndPuncttrue
\mciteSetBstMidEndSepPunct{\mcitedefaultmidpunct}
{\mcitedefaultendpunct}{\mcitedefaultseppunct}\relax
\EndOfBibitem
\bibitem[Stolyarov \latin{et~al.}(2016)Stolyarov, Liu, Bloodgood, Aytan, Jiang, Samnakay, Salguero, Nika, Rumyantsev, Shur, \latin{et~al.} others]{stolyarov2016breakdown}
Stolyarov,~M.~A.; Liu,~G.; Bloodgood,~M.~A.; Aytan,~E.; Jiang,~C.; Samnakay,~R.; Salguero,~T.~T.; Nika,~D.~L.; Rumyantsev,~S.~L.; Shur,~M.~S.; others {Breakdown current density in h-BN-capped quasi-1D TaSe3 metallic nanowires: prospects of interconnect applications}. \emph{Nanoscale} \textbf{2016}, \emph{8}, 15774--15782\relax
\mciteBstWouldAddEndPuncttrue
\mciteSetBstMidEndSepPunct{\mcitedefaultmidpunct}
{\mcitedefaultendpunct}{\mcitedefaultseppunct}\relax
\EndOfBibitem
\bibitem[Empante \latin{et~al.}(2019)Empante, Martinez, Wurch, Zhu, Geremew, Yamaguchi, Isarraraz, Rumyantsev, Reed, Balandin, \latin{et~al.} others]{empante2019low}
Empante,~T.~A.; Martinez,~A.; Wurch,~M.; Zhu,~Y.; Geremew,~A.~K.; Yamaguchi,~K.; Isarraraz,~M.; Rumyantsev,~S.; Reed,~E.~J.; Balandin,~A.~A.; others {Low resistivity and high breakdown current density of 10 nm diameter van der Waals TaSe3 nanowires by chemical vapor deposition}. \emph{Nano Letters} \textbf{2019}, \emph{19}, 4355--4361\relax
\mciteBstWouldAddEndPuncttrue
\mciteSetBstMidEndSepPunct{\mcitedefaultmidpunct}
{\mcitedefaultendpunct}{\mcitedefaultseppunct}\relax
\EndOfBibitem
\bibitem[Zhou \latin{et~al.}(2015)Zhou, Wang, Chen, Qin, Liu, Chen, Xue, Luo, Cao, Cheng, \latin{et~al.} others]{zhou2015thin}
Zhou,~Y.; Wang,~L.; Chen,~S.; Qin,~S.; Liu,~X.; Chen,~J.; Xue,~D.-J.; Luo,~M.; Cao,~Y.; Cheng,~Y.; others {Thin-film Sb2Se3 photovoltaics with oriented one-dimensional ribbons and benign grain boundaries}. \emph{Nature Photonics} \textbf{2015}, \emph{9}, 409--415\relax
\mciteBstWouldAddEndPuncttrue
\mciteSetBstMidEndSepPunct{\mcitedefaultmidpunct}
{\mcitedefaultendpunct}{\mcitedefaultseppunct}\relax
\EndOfBibitem
\bibitem[Peierls and Peierls(1955)Peierls, and Peierls]{peierls1955quantum}
Peierls,~R.; Peierls,~R.~E. \emph{Quantum theory of solids}; Oxford University Press, 1955\relax
\mciteBstWouldAddEndPuncttrue
\mciteSetBstMidEndSepPunct{\mcitedefaultmidpunct}
{\mcitedefaultendpunct}{\mcitedefaultseppunct}\relax
\EndOfBibitem
\bibitem[Luttinger(1963)]{luttinger1963exactly}
Luttinger,~J. An exactly soluble model of a many-fermion system. \emph{Journal of Mathematical Physics} \textbf{1963}, \emph{4}, 1154--1162\relax
\mciteBstWouldAddEndPuncttrue
\mciteSetBstMidEndSepPunct{\mcitedefaultmidpunct}
{\mcitedefaultendpunct}{\mcitedefaultseppunct}\relax
\EndOfBibitem
\bibitem[Varsano \latin{et~al.}(2017)Varsano, Sorella, Sangalli, Barborini, Corni, Molinari, and Rontani]{Varsano2017}
Varsano,~D.; Sorella,~S.; Sangalli,~D.; Barborini,~M.; Corni,~S.; Molinari,~E.; Rontani,~M. Carbon nanotubes as excitonic insulators. \emph{Nature Communications} \textbf{2017}, \emph{8}, 1461\relax
\mciteBstWouldAddEndPuncttrue
\mciteSetBstMidEndSepPunct{\mcitedefaultmidpunct}
{\mcitedefaultendpunct}{\mcitedefaultseppunct}\relax
\EndOfBibitem
\bibitem[Giamarchi(2016)]{giamarchi2016one}
Giamarchi,~T. One-dimensional physics in the 21st century. \emph{Comptes Rendus Physique} \textbf{2016}, \emph{17}, 322--331\relax
\mciteBstWouldAddEndPuncttrue
\mciteSetBstMidEndSepPunct{\mcitedefaultmidpunct}
{\mcitedefaultendpunct}{\mcitedefaultseppunct}\relax
\EndOfBibitem
\bibitem[Guo \latin{et~al.}(2022)Guo, Fu, Zhang, Zhu, Yao, Xu, An, Wang, Tang, Deng, \latin{et~al.} others]{guo2022direct}
Guo,~S.; Fu,~J.; Zhang,~P.; Zhu,~C.; Yao,~H.; Xu,~M.; An,~B.; Wang,~X.; Tang,~B.; Deng,~Y.; others Direct growth of single-metal-atom chains. \emph{Nature Synthesis} \textbf{2022}, \emph{1}, 245--253\relax
\mciteBstWouldAddEndPuncttrue
\mciteSetBstMidEndSepPunct{\mcitedefaultmidpunct}
{\mcitedefaultendpunct}{\mcitedefaultseppunct}\relax
\EndOfBibitem
\bibitem[Zeng \latin{et~al.}(2008)Zeng, Kent, Kim, Li, and Weitering]{zeng2008charge}
Zeng,~C.; Kent,~P.; Kim,~T.-H.; Li,~A.-P.; Weitering,~H.~H. Charge-order fluctuations in one-dimensional silicides. \emph{Nature Materials} \textbf{2008}, \emph{7}, 539--542\relax
\mciteBstWouldAddEndPuncttrue
\mciteSetBstMidEndSepPunct{\mcitedefaultmidpunct}
{\mcitedefaultendpunct}{\mcitedefaultseppunct}\relax
\EndOfBibitem
\bibitem[Gambardella \latin{et~al.}(2002)Gambardella, Dallmeyer, Maiti, Malagoli, Eberhardt, Kern, and Carbone]{gambardella2002ferromagnetism}
Gambardella,~P.; Dallmeyer,~A.; Maiti,~K.; Malagoli,~M.; Eberhardt,~W.; Kern,~K.; Carbone,~C. Ferromagnetism in one-dimensional monatomic metal chains. \emph{Nature} \textbf{2002}, \emph{416}, 301--304\relax
\mciteBstWouldAddEndPuncttrue
\mciteSetBstMidEndSepPunct{\mcitedefaultmidpunct}
{\mcitedefaultendpunct}{\mcitedefaultseppunct}\relax
\EndOfBibitem
\bibitem[Kashtiban \latin{et~al.}(2021)Kashtiban, Burdanova, Vasylenko, Wynn, Medeiros, Ramasse, Morris, Quigley, Lloyd-Hughes, and Sloan]{kashtiban2021linear}
Kashtiban,~R.~J.; Burdanova,~M.~G.; Vasylenko,~A.; Wynn,~J.; Medeiros,~P.~V.; Ramasse,~Q.; Morris,~A.~J.; Quigley,~D.; Lloyd-Hughes,~J.; Sloan,~J. Linear and Helical Cesium Iodide Atomic Chains in Ultranarrow Single-Walled Carbon Nanotubes: Impact on Optical Properties. \emph{ACS Nano} \textbf{2021}, \emph{15}, 13389--13398\relax
\mciteBstWouldAddEndPuncttrue
\mciteSetBstMidEndSepPunct{\mcitedefaultmidpunct}
{\mcitedefaultendpunct}{\mcitedefaultseppunct}\relax
\EndOfBibitem
\bibitem[Stonemeyer \latin{et~al.}(2020)Stonemeyer, Cain, Oh, Azizi, Elasha, Thiel, Song, Ercius, Cohen, and Zettl]{stonemeyer2020stabilization}
Stonemeyer,~S.; Cain,~J.~D.; Oh,~S.; Azizi,~A.; Elasha,~M.; Thiel,~M.; Song,~C.; Ercius,~P.; Cohen,~M.~L.; Zettl,~A. {Stabilization of NbTe3, VTe3, and TiTe3 via nanotube encapsulation}. \emph{Journal of the American Chemical Society} \textbf{2020}, \emph{143}, 4563--4568\relax
\mciteBstWouldAddEndPuncttrue
\mciteSetBstMidEndSepPunct{\mcitedefaultmidpunct}
{\mcitedefaultendpunct}{\mcitedefaultseppunct}\relax
\EndOfBibitem
\bibitem[Island \latin{et~al.}(2017)Island, Molina-Mendoza, Barawi, Biele, Flores, Clamagirand, Ares, S{\'a}nchez, Van Der~Zant, D’Agosta, \latin{et~al.} others]{island2017electronics}
Island,~J.~O.; Molina-Mendoza,~A.~J.; Barawi,~M.; Biele,~R.; Flores,~E.; Clamagirand,~J.~M.; Ares,~J.~R.; S{\'a}nchez,~C.; Van Der~Zant,~H.~S.; D’Agosta,~R.; others {Electronics and optoelectronics of quasi-1D layered transition metal trichalcogenides}. \emph{2D Materials} \textbf{2017}, \emph{4}, 022003\relax
\mciteBstWouldAddEndPuncttrue
\mciteSetBstMidEndSepPunct{\mcitedefaultmidpunct}
{\mcitedefaultendpunct}{\mcitedefaultseppunct}\relax
\EndOfBibitem
\bibitem[Larsen \latin{et~al.}(2019)Larsen, Pandey, Strange, and Jacobsen]{larsen2019definition}
Larsen,~P.~M.; Pandey,~M.; Strange,~M.; Jacobsen,~K.~W. Definition of a scoring parameter to identify low-dimensional materials components. \emph{Physical Review Materials} \textbf{2019}, \emph{3}, 034003\relax
\mciteBstWouldAddEndPuncttrue
\mciteSetBstMidEndSepPunct{\mcitedefaultmidpunct}
{\mcitedefaultendpunct}{\mcitedefaultseppunct}\relax
\EndOfBibitem
\bibitem[Moustafa \latin{et~al.}(2022)Moustafa, Larsen, Gjerding, Mortensen, Thygesen, and Jacobsen]{moustafa2022computational}
Moustafa,~H.; Larsen,~P.~M.; Gjerding,~M.~N.; Mortensen,~J.~J.; Thygesen,~K.~S.; Jacobsen,~K.~W. {Computational exfoliation of atomically thin one-dimensional materials with application to Majorana bound states}. \emph{Physical Review Materials} \textbf{2022}, \emph{6}, 064202\relax
\mciteBstWouldAddEndPuncttrue
\mciteSetBstMidEndSepPunct{\mcitedefaultmidpunct}
{\mcitedefaultendpunct}{\mcitedefaultseppunct}\relax
\EndOfBibitem
\bibitem[Campi \latin{et~al.}()Campi, Cignarella, and Marzari]{campiinpreparation}
Campi,~D.; Cignarella,~C.; Marzari,~N. High-throughput screening for one-dimensional exfoliable materials. \emph{In preparation} \relax
\mciteBstWouldAddEndPunctfalse
\mciteSetBstMidEndSepPunct{\mcitedefaultmidpunct}
{}{\mcitedefaultseppunct}\relax
\EndOfBibitem
\bibitem[Cignarella \latin{et~al.}(2024)Cignarella, Campi, and Marzari]{cignarella2024searching}
Cignarella,~C.; Campi,~D.; Marzari,~N. Searching for the thinnest metallic wire. \emph{ACS nano} \textbf{2024}, \emph{18}, 16101--16112\relax
\mciteBstWouldAddEndPuncttrue
\mciteSetBstMidEndSepPunct{\mcitedefaultmidpunct}
{\mcitedefaultendpunct}{\mcitedefaultseppunct}\relax
\EndOfBibitem
\bibitem[Cheon \latin{et~al.}(2017)Cheon, Duerloo, Sendek, Porter, Chen, and Reed]{cheon2017data}
Cheon,~G.; Duerloo,~K.-A.~N.; Sendek,~A.~D.; Porter,~C.; Chen,~Y.; Reed,~E.~J. Data mining for new two-and one-dimensional weakly bonded solids and lattice-commensurate heterostructures. \emph{Nano Letters} \textbf{2017}, \emph{17}, 1915--1923\relax
\mciteBstWouldAddEndPuncttrue
\mciteSetBstMidEndSepPunct{\mcitedefaultmidpunct}
{\mcitedefaultendpunct}{\mcitedefaultseppunct}\relax
\EndOfBibitem
\bibitem[Mounet \latin{et~al.}(2018)Mounet, Gibertini, Schwaller, Campi, Merkys, Marrazzo, Sohier, Castelli, Cepellotti, Pizzi, \latin{et~al.} others]{mounet2018two}
Mounet,~N.; Gibertini,~M.; Schwaller,~P.; Campi,~D.; Merkys,~A.; Marrazzo,~A.; Sohier,~T.; Castelli,~I.~E.; Cepellotti,~A.; Pizzi,~G.; others Two-dimensional materials from high-throughput computational exfoliation of experimentally known compounds. \emph{Nature Nanotechnology} \textbf{2018}, \emph{13}, 246--252\relax
\mciteBstWouldAddEndPuncttrue
\mciteSetBstMidEndSepPunct{\mcitedefaultmidpunct}
{\mcitedefaultendpunct}{\mcitedefaultseppunct}\relax
\EndOfBibitem
\bibitem[Haastrup \latin{et~al.}(2018)Haastrup, Strange, Pandey, Deilmann, Schmidt, Hinsche, Gjerding, Torelli, Larsen, Riis-Jensen, \latin{et~al.} others]{haastrup2018computational}
Haastrup,~S.; Strange,~M.; Pandey,~M.; Deilmann,~T.; Schmidt,~P.~S.; Hinsche,~N.~F.; Gjerding,~M.~N.; Torelli,~D.; Larsen,~P.~M.; Riis-Jensen,~A.~C.; others {The Computational 2D Materials Database: high-throughput modeling and discovery of atomically thin crystals}. \emph{2D Materials} \textbf{2018}, \emph{5}, 042002\relax
\mciteBstWouldAddEndPuncttrue
\mciteSetBstMidEndSepPunct{\mcitedefaultmidpunct}
{\mcitedefaultendpunct}{\mcitedefaultseppunct}\relax
\EndOfBibitem
\bibitem[Campi \latin{et~al.}(2023)Campi, Mounet, Gibertini, Pizzi, and Marzari]{campi2023expansion}
Campi,~D.; Mounet,~N.; Gibertini,~M.; Pizzi,~G.; Marzari,~N. Expansion of the Materials Cloud 2D Database. \emph{ACS Nano} \textbf{2023}, \emph{17}, 11268--11278\relax
\mciteBstWouldAddEndPuncttrue
\mciteSetBstMidEndSepPunct{\mcitedefaultmidpunct}
{\mcitedefaultendpunct}{\mcitedefaultseppunct}\relax
\EndOfBibitem
\bibitem[Liu \latin{et~al.}(2017)Liu, Rumyantsev, Bloodgood, Salguero, Shur, and Balandin]{liu2017low}
Liu,~G.; Rumyantsev,~S.; Bloodgood,~M.~A.; Salguero,~T.~T.; Shur,~M.; Balandin,~A.~A. {Low-frequency electronic noise in quasi-1D TaSe3 van der Waals nanowires}. \emph{Nano Letters} \textbf{2017}, \emph{17}, 377--383\relax
\mciteBstWouldAddEndPuncttrue
\mciteSetBstMidEndSepPunct{\mcitedefaultmidpunct}
{\mcitedefaultendpunct}{\mcitedefaultseppunct}\relax
\EndOfBibitem
\bibitem[Liu \latin{et~al.}(2016)Liu, Debnath, Pope, Salguero, Lake, and Balandin]{liu2016charge}
Liu,~G.; Debnath,~B.; Pope,~T.~R.; Salguero,~T.~T.; Lake,~R.~K.; Balandin,~A.~A. A charge-density-wave oscillator based on an integrated tantalum disulfide--boron nitride--graphene device operating at room temperature. \emph{Nature Nanotechnology} \textbf{2016}, \emph{11}, 845--850\relax
\mciteBstWouldAddEndPuncttrue
\mciteSetBstMidEndSepPunct{\mcitedefaultmidpunct}
{\mcitedefaultendpunct}{\mcitedefaultseppunct}\relax
\EndOfBibitem
\bibitem[Zhu \latin{et~al.}(2018)Zhu, Chen, Liu, Zheng, Li, Chaturvedi, Zhou, Fu, He, Zeng, \latin{et~al.} others]{zhu2018light}
Zhu,~C.; Chen,~Y.; Liu,~F.; Zheng,~S.; Li,~X.; Chaturvedi,~A.; Zhou,~J.; Fu,~Q.; He,~Y.; Zeng,~Q.; others {Light-tunable 1T-TaS2 charge-density-wave oscillators}. \emph{ACS Nano} \textbf{2018}, \emph{12}, 11203--11210\relax
\mciteBstWouldAddEndPuncttrue
\mciteSetBstMidEndSepPunct{\mcitedefaultmidpunct}
{\mcitedefaultendpunct}{\mcitedefaultseppunct}\relax
\EndOfBibitem
\bibitem[Geremew \latin{et~al.}(2019)Geremew, Rumyantsev, Kargar, Debnath, Nosek, Bloodgood, Bockrath, Salguero, Lake, and Balandin]{geremew2019bias}
Geremew,~A.~K.; Rumyantsev,~S.; Kargar,~F.; Debnath,~B.; Nosek,~A.; Bloodgood,~M.~A.; Bockrath,~M.; Salguero,~T.~T.; Lake,~R.~K.; Balandin,~A.~A. {Bias-voltage driven switching of the charge-density-wave and normal metallic phases in 1T-TaS2 thin-film devices}. \emph{ACS Nano} \textbf{2019}, \emph{13}, 7231--7240\relax
\mciteBstWouldAddEndPuncttrue
\mciteSetBstMidEndSepPunct{\mcitedefaultmidpunct}
{\mcitedefaultendpunct}{\mcitedefaultseppunct}\relax
\EndOfBibitem
\bibitem[Ceperley(1995)]{ceperley1995path}
Ceperley,~D.~M. Path integrals in the theory of condensed helium. \emph{Reviews of Modern Physics} \textbf{1995}, \emph{67}, 279\relax
\mciteBstWouldAddEndPuncttrue
\mciteSetBstMidEndSepPunct{\mcitedefaultmidpunct}
{\mcitedefaultendpunct}{\mcitedefaultseppunct}\relax
\EndOfBibitem
\bibitem[Errea \latin{et~al.}(2014)Errea, Calandra, and Mauri]{errea2014anharmonic}
Errea,~I.; Calandra,~M.; Mauri,~F. Anharmonic free energies and phonon dispersions from the stochastic self-consistent harmonic approximation: application to platinum and palladium hydrides. \emph{Physical Review B} \textbf{2014}, \emph{89}, 064302\relax
\mciteBstWouldAddEndPuncttrue
\mciteSetBstMidEndSepPunct{\mcitedefaultmidpunct}
{\mcitedefaultendpunct}{\mcitedefaultseppunct}\relax
\EndOfBibitem
\bibitem[Monacelli \latin{et~al.}(2021)Monacelli, Bianco, Cherubini, Calandra, Errea, and Mauri]{monacelli2021stochastic}
Monacelli,~L.; Bianco,~R.; Cherubini,~M.; Calandra,~M.; Errea,~I.; Mauri,~F. The stochastic self-consistent harmonic approximation: calculating vibrational properties of materials with full quantum and anharmonic effects. \emph{Journal of Physics: Condensed Matter} \textbf{2021}, \emph{33}, 363001\relax
\mciteBstWouldAddEndPuncttrue
\mciteSetBstMidEndSepPunct{\mcitedefaultmidpunct}
{\mcitedefaultendpunct}{\mcitedefaultseppunct}\relax
\EndOfBibitem
\bibitem[Errea \latin{et~al.}(2020)Errea, Belli, Monacelli, Sanna, Koretsune, Tadano, Bianco, Calandra, Arita, Mauri, and Flores-Livas]{errea_quantum_2020}
Errea,~I.; Belli,~F.; Monacelli,~L.; Sanna,~A.; Koretsune,~T.; Tadano,~T.; Bianco,~R.; Calandra,~M.; Arita,~R.; Mauri,~F.; Flores-Livas,~J.~A. Quantum crystal structure in the 250-kelvin superconducting lanthanum hydride. \emph{Nature} \textbf{2020}, 66--69\relax
\mciteBstWouldAddEndPuncttrue
\mciteSetBstMidEndSepPunct{\mcitedefaultmidpunct}
{\mcitedefaultendpunct}{\mcitedefaultseppunct}\relax
\EndOfBibitem
\bibitem[Romanin \latin{et~al.}(2021)Romanin, Monacelli, Bianco, Errea, Mauri, and Calandra]{romanin2021dominant}
Romanin,~D.; Monacelli,~L.; Bianco,~R.; Errea,~I.; Mauri,~F.; Calandra,~M. Dominant role of quantum anharmonicity in the stability and optical properties of infinite linear acetylenic carbon chains. \emph{The Journal of Physical Chemistry Letters} \textbf{2021}, \emph{12}, 10339--10345\relax
\mciteBstWouldAddEndPuncttrue
\mciteSetBstMidEndSepPunct{\mcitedefaultmidpunct}
{\mcitedefaultendpunct}{\mcitedefaultseppunct}\relax
\EndOfBibitem
\bibitem[Monacelli \latin{et~al.}(2023)Monacelli, Casula, Nakano, Sorella, and Mauri]{monacelli_quantum_2023}
Monacelli,~L.; Casula,~M.; Nakano,~K.; Sorella,~S.; Mauri,~F. Quantum phase diagram of high-pressure hydrogen. \emph{Nature Physics} \textbf{2023}, \emph{19}, 845--850\relax
\mciteBstWouldAddEndPuncttrue
\mciteSetBstMidEndSepPunct{\mcitedefaultmidpunct}
{\mcitedefaultendpunct}{\mcitedefaultseppunct}\relax
\EndOfBibitem
\bibitem[Monacelli and Marzari(2023)Monacelli, and Marzari]{monacelli2023first}
Monacelli,~L.; Marzari,~N. {First-principles thermodynamics of CsSnI3}. \emph{Chemistry of Materials} \textbf{2023}, \emph{35}, 1702--1709\relax
\mciteBstWouldAddEndPuncttrue
\mciteSetBstMidEndSepPunct{\mcitedefaultmidpunct}
{\mcitedefaultendpunct}{\mcitedefaultseppunct}\relax
\EndOfBibitem
\bibitem[Monacelli(2025)]{monacelli_simulating_2024}
Monacelli,~L. Analyzing the anharmonic phonon spectrum: Self-consistent approximation and temperature-dependent effective potential methods. \emph{Physical Review B} \textbf{2025}, \emph{112}, 014109\relax
\mciteBstWouldAddEndPuncttrue
\mciteSetBstMidEndSepPunct{\mcitedefaultmidpunct}
{\mcitedefaultendpunct}{\mcitedefaultseppunct}\relax
\EndOfBibitem
\bibitem[Lin \latin{et~al.}(2022)Lin, Ponc{\'e}, and Marzari]{lin2022general}
Lin,~C.; Ponc{\'e},~S.; Marzari,~N. {General invariance and equilibrium conditions for lattice dynamics in 1D, 2D, and 3D materials}. \emph{npj Computational Materials} \textbf{2022}, \emph{8}, 236\relax
\mciteBstWouldAddEndPuncttrue
\mciteSetBstMidEndSepPunct{\mcitedefaultmidpunct}
{\mcitedefaultendpunct}{\mcitedefaultseppunct}\relax
\EndOfBibitem
\bibitem[SSC()]{SSCHA_1d}
Branch {asr\_1d} SSCHA, \textit{GitHub repository}. \url{https://github.com/SSCHAcode/python-sscha/tree/asr_1d}, Accessed: 11-08-2025\relax
\mciteBstWouldAddEndPuncttrue
\mciteSetBstMidEndSepPunct{\mcitedefaultmidpunct}
{\mcitedefaultendpunct}{\mcitedefaultseppunct}\relax
\EndOfBibitem
\bibitem[Aseginolaza \latin{et~al.}(2024)Aseginolaza, Diego, Cea, Bianco, Monacelli, Libbi, Calandra, Bergara, Mauri, and Errea]{aseginolaza2024bending}
Aseginolaza,~U.; Diego,~J.; Cea,~T.; Bianco,~R.; Monacelli,~L.; Libbi,~F.; Calandra,~M.; Bergara,~A.; Mauri,~F.; Errea,~I. Bending rigidity, sound propagation and ripples in flat graphene. \emph{Nature Physics} \textbf{2024}, 1--6\relax
\mciteBstWouldAddEndPuncttrue
\mciteSetBstMidEndSepPunct{\mcitedefaultmidpunct}
{\mcitedefaultendpunct}{\mcitedefaultseppunct}\relax
\EndOfBibitem
\bibitem[Batzner \latin{et~al.}(2022)Batzner, Musaelian, Sun, Geiger, Mailoa, Kornbluth, Molinari, Smidt, and Kozinsky]{batzner20223}
Batzner,~S.; Musaelian,~A.; Sun,~L.; Geiger,~M.; Mailoa,~J.~P.; Kornbluth,~M.; Molinari,~N.; Smidt,~T.~E.; Kozinsky,~B. E(3)-equivariant graph neural networks for data-efficient and accurate interatomic potentials. \emph{Nature Communications} \textbf{2022}, \emph{13}, 2453\relax
\mciteBstWouldAddEndPuncttrue
\mciteSetBstMidEndSepPunct{\mcitedefaultmidpunct}
{\mcitedefaultendpunct}{\mcitedefaultseppunct}\relax
\EndOfBibitem
\bibitem[Mounet and Marzari(2005)Mounet, and Marzari]{mounet2005first}
Mounet,~N.; Marzari,~N. First-principles determination of the structural, vibrational and thermodynamic properties of diamond, graphite, and derivatives. \emph{Physical Review B} \textbf{2005}, \emph{71}, 205214\relax
\mciteBstWouldAddEndPuncttrue
\mciteSetBstMidEndSepPunct{\mcitedefaultmidpunct}
{\mcitedefaultendpunct}{\mcitedefaultseppunct}\relax
\EndOfBibitem
\bibitem[Kwon \latin{et~al.}(2004)Kwon, Berber, and Tom{\'a}nek]{kwon2004thermal}
Kwon,~Y.-K.; Berber,~S.; Tom{\'a}nek,~D. Thermal contraction of carbon fullerenes and nanotubes. \emph{Physical Review Letters} \textbf{2004}, \emph{92}, 015901\relax
\mciteBstWouldAddEndPuncttrue
\mciteSetBstMidEndSepPunct{\mcitedefaultmidpunct}
{\mcitedefaultendpunct}{\mcitedefaultseppunct}\relax
\EndOfBibitem
\bibitem[Ashcroft and Mermin(1976)Ashcroft, and Mermin]{ashcroft1976solid}
Ashcroft,~N.~W.; Mermin,~N. \emph{Solid state}; Holt, Rinehart and Winston, 1976; Chapter 25\relax
\mciteBstWouldAddEndPuncttrue
\mciteSetBstMidEndSepPunct{\mcitedefaultmidpunct}
{\mcitedefaultendpunct}{\mcitedefaultseppunct}\relax
\EndOfBibitem
\bibitem[Br{\"u}esch(2012)]{bruesch2012phonons}
Br{\"u}esch,~P. \emph{Phonons: Theory and experiments I: Lattice dynamics and Models of interatomic forces}; Springer Science \& Business Media, 2012; Vol.~34; Chapter 5\relax
\mciteBstWouldAddEndPuncttrue
\mciteSetBstMidEndSepPunct{\mcitedefaultmidpunct}
{\mcitedefaultendpunct}{\mcitedefaultseppunct}\relax
\EndOfBibitem
\bibitem[Libbi \latin{et~al.}(2020)Libbi, Bonini, and Marzari]{libbi2020thermomechanical}
Libbi,~F.; Bonini,~N.; Marzari,~N. Thermomechanical properties of honeycomb lattices from internal-coordinates potentials: the case of graphene and hexagonal boron nitride. \emph{2D Materials} \textbf{2020}, \emph{8}, 015026\relax
\mciteBstWouldAddEndPuncttrue
\mciteSetBstMidEndSepPunct{\mcitedefaultmidpunct}
{\mcitedefaultendpunct}{\mcitedefaultseppunct}\relax
\EndOfBibitem
\bibitem[Bianco \latin{et~al.}(2017)Bianco, Errea, Paulatto, Calandra, and Mauri]{bianco_2017_structural}
Bianco,~R.; Errea,~I.; Paulatto,~L.; Calandra,~M.; Mauri,~F. Second-order structural phase transitions, free energy curvature, and temperature-dependent anharmonic phonons in the self-consistent harmonic approximation: Theory and stochastic implementation. \emph{Physical Review B} \textbf{2017}, \emph{96}, 014111\relax
\mciteBstWouldAddEndPuncttrue
\mciteSetBstMidEndSepPunct{\mcitedefaultmidpunct}
{\mcitedefaultendpunct}{\mcitedefaultseppunct}\relax
\EndOfBibitem
\bibitem[Miller \latin{et~al.}(2009)Miller, Smith, Mackenzie, and Evans]{miller2009negative}
Miller,~W.; Smith,~C.; Mackenzie,~D.; Evans,~K. Negative thermal expansion: a review. \emph{Journal of materials science} \textbf{2009}, \emph{44}, 5441--5451\relax
\mciteBstWouldAddEndPuncttrue
\mciteSetBstMidEndSepPunct{\mcitedefaultmidpunct}
{\mcitedefaultendpunct}{\mcitedefaultseppunct}\relax
\EndOfBibitem
\bibitem[Eremenko \latin{et~al.}(2016)Eremenko, Sirenko, Sirenko, Dolbin, Gospodarev, Syrkin, Feodosyev, Bondar, and Minakova]{eremenko2016role}
Eremenko,~V.; Sirenko,~A.; Sirenko,~V.; Dolbin,~A.; Gospodarev,~I.; Syrkin,~E.; Feodosyev,~S.; Bondar,~I.; Minakova,~K. Role of acoustic phonons in the negative thermal expansion of layered structures and nanotubes based on them. \emph{Low Temperature Physics} \textbf{2016}, \emph{42}, 401--410\relax
\mciteBstWouldAddEndPuncttrue
\mciteSetBstMidEndSepPunct{\mcitedefaultmidpunct}
{\mcitedefaultendpunct}{\mcitedefaultseppunct}\relax
\EndOfBibitem
\bibitem[Barrera \latin{et~al.}(2005)Barrera, Bruno, Barron, and Allan]{barrera2005negative}
Barrera,~G.~D.; Bruno,~J. A.~O.; Barron,~T.; Allan,~N. Negative thermal expansion. \emph{Journal of Physics: Condensed Matter} \textbf{2005}, \emph{17}, R217\relax
\mciteBstWouldAddEndPuncttrue
\mciteSetBstMidEndSepPunct{\mcitedefaultmidpunct}
{\mcitedefaultendpunct}{\mcitedefaultseppunct}\relax
\EndOfBibitem
\bibitem[Schelling and Keblinski(2003)Schelling, and Keblinski]{schelling2003thermal}
Schelling,~P.; Keblinski,~P. Thermal expansion of carbon structures. \emph{Physical Review B} \textbf{2003}, \emph{68}, 035425\relax
\mciteBstWouldAddEndPuncttrue
\mciteSetBstMidEndSepPunct{\mcitedefaultmidpunct}
{\mcitedefaultendpunct}{\mcitedefaultseppunct}\relax
\EndOfBibitem
\bibitem[Ho \latin{et~al.}(2017)Ho, Kwon, Park, and Kim]{ho2017negative}
Ho,~D.~T.; Kwon,~S.-Y.; Park,~H.~S.; Kim,~S.~Y. Negative thermal expansion of ultrathin metal nanowires: a computational study. \emph{Nano letters} \textbf{2017}, \emph{17}, 5113--5118\relax
\mciteBstWouldAddEndPuncttrue
\mciteSetBstMidEndSepPunct{\mcitedefaultmidpunct}
{\mcitedefaultendpunct}{\mcitedefaultseppunct}\relax
\EndOfBibitem
\bibitem[Jiang \latin{et~al.}(2010)Jiang, Wang, and Li]{jiang2010thermal}
Jiang,~J.-W.; Wang,~J.-S.; Li,~B. Thermal contraction in silicon nanowires at low temperatures. \emph{Nanoscale} \textbf{2010}, \emph{2}, 2864--2867\relax
\mciteBstWouldAddEndPuncttrue
\mciteSetBstMidEndSepPunct{\mcitedefaultmidpunct}
{\mcitedefaultendpunct}{\mcitedefaultseppunct}\relax
\EndOfBibitem
\bibitem[Yoon \latin{et~al.}(2011)Yoon, Son, and Cheong]{yoon2011negative}
Yoon,~D.; Son,~Y.-W.; Cheong,~H. Negative thermal expansion coefficient of graphene measured by {Raman} spectroscopy. \emph{Nano Letters} \textbf{2011}, \emph{11}, 3227--3231\relax
\mciteBstWouldAddEndPuncttrue
\mciteSetBstMidEndSepPunct{\mcitedefaultmidpunct}
{\mcitedefaultendpunct}{\mcitedefaultseppunct}\relax
\EndOfBibitem
\bibitem[Giustino(2017)]{giustino2017electron}
Giustino,~F. Electron-phonon interactions from first principles. \emph{Reviews of Modern Physics} \textbf{2017}, \emph{89}, 015003\relax
\mciteBstWouldAddEndPuncttrue
\mciteSetBstMidEndSepPunct{\mcitedefaultmidpunct}
{\mcitedefaultendpunct}{\mcitedefaultseppunct}\relax
\EndOfBibitem
\bibitem[Monacelli \latin{et~al.}(2021)Monacelli, Errea, Calandra, and Mauri]{monacelli2021black}
Monacelli,~L.; Errea,~I.; Calandra,~M.; Mauri,~F. Black metal hydrogen above 360 GPa driven by proton quantum fluctuations. \emph{Nature Physics} \textbf{2021}, \emph{17}, 63--67\relax
\mciteBstWouldAddEndPuncttrue
\mciteSetBstMidEndSepPunct{\mcitedefaultmidpunct}
{\mcitedefaultendpunct}{\mcitedefaultseppunct}\relax
\EndOfBibitem
\bibitem[Wu \latin{et~al.}(2003)Wu, Walukiewicz, Shan, Yu, Ager~Iii, Li, Haller, Lu, and Schaff]{wu2003temperature}
Wu,~J.; Walukiewicz,~W.; Shan,~W.; Yu,~K.; Ager~Iii,~J.; Li,~S.; Haller,~E.; Lu,~H.; Schaff,~W.~J. Temperature dependence of the fundamental band gap of InN. \emph{Journal of Applied Physics} \textbf{2003}, \emph{94}, 4457--4460\relax
\mciteBstWouldAddEndPuncttrue
\mciteSetBstMidEndSepPunct{\mcitedefaultmidpunct}
{\mcitedefaultendpunct}{\mcitedefaultseppunct}\relax
\EndOfBibitem
\bibitem[Choi \latin{et~al.}(2017)Choi, Kim, Jung, Kim, Yu, and Chang]{choi2017temperature}
Choi,~B.~K.; Kim,~M.; Jung,~K.-H.; Kim,~J.; Yu,~K.-S.; Chang,~Y.~J. Temperature dependence of band gap in MoSe2 grown by molecular beam epitaxy. \emph{Nanoscale research letters} \textbf{2017}, \emph{12}, 1--7\relax
\mciteBstWouldAddEndPuncttrue
\mciteSetBstMidEndSepPunct{\mcitedefaultmidpunct}
{\mcitedefaultendpunct}{\mcitedefaultseppunct}\relax
\EndOfBibitem
\bibitem[Jiang \latin{et~al.}(2004)Jiang, Liu, Huang, and Hwang]{jiang2004thermal}
Jiang,~H.; Liu,~B.; Huang,~Y.; Hwang,~K. Thermal expansion of single wall carbon nanotubes. \emph{J. Eng. Mater. Technol.} \textbf{2004}, \emph{126}, 265--270\relax
\mciteBstWouldAddEndPuncttrue
\mciteSetBstMidEndSepPunct{\mcitedefaultmidpunct}
{\mcitedefaultendpunct}{\mcitedefaultseppunct}\relax
\EndOfBibitem
\bibitem[Bao \latin{et~al.}(2009)Bao, Miao, Chen, Zhang, Jang, Dames, and Lau]{bao2009controlled}
Bao,~W.; Miao,~F.; Chen,~Z.; Zhang,~H.; Jang,~W.; Dames,~C.; Lau,~C.~N. Controlled ripple texturing of suspended graphene and ultrathin graphite membranes. \emph{Nature nanotechnology} \textbf{2009}, \emph{4}, 562--566\relax
\mciteBstWouldAddEndPuncttrue
\mciteSetBstMidEndSepPunct{\mcitedefaultmidpunct}
{\mcitedefaultendpunct}{\mcitedefaultseppunct}\relax
\EndOfBibitem
\bibitem[Kriegel \latin{et~al.}(2023)Kriegel, Omambac, Franzka, zu~Heringdorf, and Horn-von Hoegen]{kriegel2023incommensurability}
Kriegel,~M.~A.; Omambac,~K.~M.; Franzka,~S.; zu~Heringdorf,~F.-J.~M.; Horn-von Hoegen,~M. Incommensurability and negative thermal expansion of single layer hexagonal boron nitride. \emph{Applied Surface Science} \textbf{2023}, \emph{624}, 157156\relax
\mciteBstWouldAddEndPuncttrue
\mciteSetBstMidEndSepPunct{\mcitedefaultmidpunct}
{\mcitedefaultendpunct}{\mcitedefaultseppunct}\relax
\EndOfBibitem
\bibitem[Demiroglu and Sevik(2021)Demiroglu, and Sevik]{demiroglu2021extraordinary}
Demiroglu,~I.; Sevik,~C. Extraordinary negative thermal expansion of two-dimensional nitrides: A comparative ab initio study of quasiharmonic approximation and molecular dynamics simulations. \emph{Physical Review B} \textbf{2021}, \emph{103}, 085430\relax
\mciteBstWouldAddEndPuncttrue
\mciteSetBstMidEndSepPunct{\mcitedefaultmidpunct}
{\mcitedefaultendpunct}{\mcitedefaultseppunct}\relax
\EndOfBibitem
\bibitem[Ernst \latin{et~al.}(1998)Ernst, Broholm, Kowach, and Ramirez]{ernst1998phonon}
Ernst,~G.; Broholm,~C.; Kowach,~G.; Ramirez,~A. Phonon density of states and negative thermal expansion in ZrW2O8. \emph{Nature} \textbf{1998}, \emph{396}, 147--149\relax
\mciteBstWouldAddEndPuncttrue
\mciteSetBstMidEndSepPunct{\mcitedefaultmidpunct}
{\mcitedefaultendpunct}{\mcitedefaultseppunct}\relax
\EndOfBibitem
\bibitem[Mary \latin{et~al.}(1996)Mary, Evans, Vogt, and Sleight]{mary1996negative}
Mary,~T.~A.; Evans,~J.; Vogt,~T.; Sleight,~A. Negative thermal expansion from 0.3 to 1050 Kelvin in ZrW2O8. \emph{Science} \textbf{1996}, \emph{272}, 90--92\relax
\mciteBstWouldAddEndPuncttrue
\mciteSetBstMidEndSepPunct{\mcitedefaultmidpunct}
{\mcitedefaultendpunct}{\mcitedefaultseppunct}\relax
\EndOfBibitem
\bibitem[Sanson \latin{et~al.}(2006)Sanson, Rocca, Dalba, Fornasini, Grisenti, Dapiaggi, and Artioli]{sanson2006negative}
Sanson,~A.; Rocca,~F.; Dalba,~G.; Fornasini,~P.; Grisenti,~R.; Dapiaggi,~M.; Artioli,~G. Negative thermal expansion and local dynamics in Cu 2 O and Ag 2 O. \emph{Physical Review B—Condensed Matter and Materials Physics} \textbf{2006}, \emph{73}, 214305\relax
\mciteBstWouldAddEndPuncttrue
\mciteSetBstMidEndSepPunct{\mcitedefaultmidpunct}
{\mcitedefaultendpunct}{\mcitedefaultseppunct}\relax
\EndOfBibitem
\bibitem[Greve \latin{et~al.}(2010)Greve, Martin, Lee, Chupas, Chapman, and Wilkinson]{greve2010pronounced}
Greve,~B.~K.; Martin,~K.~L.; Lee,~P.~L.; Chupas,~P.~J.; Chapman,~K.~W.; Wilkinson,~A.~P. Pronounced negative thermal expansion from a simple structure: cubic ScF3. \emph{Journal of the American Chemical Society} \textbf{2010}, \emph{132}, 15496--15498\relax
\mciteBstWouldAddEndPuncttrue
\mciteSetBstMidEndSepPunct{\mcitedefaultmidpunct}
{\mcitedefaultendpunct}{\mcitedefaultseppunct}\relax
\EndOfBibitem
\bibitem[Goodwin \latin{et~al.}(2008)Goodwin, Calleja, Conterio, Dove, Evans, Keen, Peters, and Tucker]{goodwin2008colossal}
Goodwin,~A.~L.; Calleja,~M.; Conterio,~M.~J.; Dove,~M.~T.; Evans,~J.~S.; Keen,~D.~A.; Peters,~L.; Tucker,~M.~G. Colossal positive and negative thermal expansion in the framework material Ag3 [Co (CN) 6]. \emph{Science} \textbf{2008}, \emph{319}, 794--797\relax
\mciteBstWouldAddEndPuncttrue
\mciteSetBstMidEndSepPunct{\mcitedefaultmidpunct}
{\mcitedefaultendpunct}{\mcitedefaultseppunct}\relax
\EndOfBibitem
\bibitem[R{\"o}ttger \latin{et~al.}(1994)R{\"o}ttger, Endriss, Ihringer, Doyle, and Kuhs]{rottger1994lattice}
R{\"o}ttger,~K.; Endriss,~A.; Ihringer,~J.; Doyle,~S.; Kuhs,~W. Lattice constants and thermal expansion of H2O and D2O ice Ih between 10 and 265 K. \emph{Acta Crystallographica Section B: Structural Science} \textbf{1994}, \emph{50}, 644--648\relax
\mciteBstWouldAddEndPuncttrue
\mciteSetBstMidEndSepPunct{\mcitedefaultmidpunct}
{\mcitedefaultendpunct}{\mcitedefaultseppunct}\relax
\EndOfBibitem
\bibitem[Fortes(2018)]{fortes2018accurate}
Fortes,~A.~D. Accurate and precise lattice parameters of H2O and D2O ice Ih between 1.6 and 270 K from high-resolution time-of-flight neutron powder diffraction data. \emph{Acta Crystallographica Section B: Structural Science, Crystal Engineering and Materials} \textbf{2018}, \emph{74}, 196--216\relax
\mciteBstWouldAddEndPuncttrue
\mciteSetBstMidEndSepPunct{\mcitedefaultmidpunct}
{\mcitedefaultendpunct}{\mcitedefaultseppunct}\relax
\EndOfBibitem
\bibitem[Chapman \latin{et~al.}(2005)Chapman, Chupas, and Kepert]{chapman2005direct}
Chapman,~K.~W.; Chupas,~P.~J.; Kepert,~C.~J. Direct observation of a transverse vibrational mechanism for negative thermal expansion in Zn (CN) 2: an atomic pair distribution function analysis. \emph{Journal of the American Chemical Society} \textbf{2005}, \emph{127}, 15630--15636\relax
\mciteBstWouldAddEndPuncttrue
\mciteSetBstMidEndSepPunct{\mcitedefaultmidpunct}
{\mcitedefaultendpunct}{\mcitedefaultseppunct}\relax
\EndOfBibitem
\bibitem[Lifshitz(1952)]{lifshitz1952thermal}
Lifshitz,~I.~M. Thermal properties of chain and layered structures at low temperatures. \emph{Zh. Eksp. Teor. Fiz} \textbf{1952}, \emph{22}, 475--486\relax
\mciteBstWouldAddEndPuncttrue
\mciteSetBstMidEndSepPunct{\mcitedefaultmidpunct}
{\mcitedefaultendpunct}{\mcitedefaultseppunct}\relax
\EndOfBibitem
\bibitem[SSC()]{SSCHA_github}
SSCHA \textit{GitHub repository}. \url{https://github.com/SSCHAcode}, Accessed: 11-08-2025\relax
\mciteBstWouldAddEndPuncttrue
\mciteSetBstMidEndSepPunct{\mcitedefaultmidpunct}
{\mcitedefaultendpunct}{\mcitedefaultseppunct}\relax
\EndOfBibitem
\bibitem[Cignarella \latin{et~al.}(2025)Cignarella, Bastonero, Monacelli, and Marzari]{cignarella2025cloud}
Cignarella,~C.; Bastonero,~L.; Monacelli,~L.; Marzari,~N. Extreme anharmonicity and thermal contraction of 1D wires. \emph{Materials Cloud Archive} \textbf{2025}, 2025.149, DOI: \href{https://doi.org/10.24435/materialscloud:fj-20}{10.24435/materialscloud:fj-20}\relax
\mciteBstWouldAddEndPuncttrue
\mciteSetBstMidEndSepPunct{\mcitedefaultmidpunct}
{\mcitedefaultendpunct}{\mcitedefaultseppunct}\relax
\EndOfBibitem
\end{mcitethebibliography}

\end{document}